# Data Driven Optimizations for MTJ based Stochastic Computing

Ankit Mondal and Ankur Srivastava
Department of Electrical and Computer Engineering, University of Maryland, College Park, MD 20742, USA

**Abstract** –Stochastic computing, a form of computation with probabilities, presents an alternative to conventional arithmetic units. Magnetic Tunnel Junctions (MTJs), which exhibit probabilistic switching, have been explored as Stochastic Number Generators (SNGs). We provide a perspective of the energy requirements of such an application and design an energy-efficient and data-sensitive MTJ-based SNG. We discuss its benefits when used for stochastic computations, illustrating with the help of a multiplier circuit, in terms of energy savings when compared to computing with the baseline MTJ-SNG.

## INTRODUCTION

Approximate Computing refers to the computation of imprecise results for certain categories of applications where an exact result is not needed in order to save energy [1]. It is thus mainly about the trade-off between the quality of results and the energy spent in computing it. Stochastic Computing (SC), on similar lines, concerns with the use of low-cost arithmetic units instead of conventional binary circuits [2]. It aims to exploit the probabilistic nature of certain computations arising due to manufacturing process variations or soft errors in modern circuits. In SC, data, which are interpreted as probabilities and called stochastic numbers (SNs), are represented in the form of bit streams of 0s and 1s and generated by circuits called stochastic number generators (SNGs). The most common SNG is a combination of a Linear Feedback Shift Register (LFSR) and a comparator.

MTJs exhibit spin-transfer effects – spin polarized current, when passed through an MTJ, can switch the magnetization of its free layer. The Spin-transfer Torque RAM, which is based on MTJs, has been explored as a memory device and is particularly attractive due to its non-volatility [3]. While a lot of research has focused on reducing the critical switching current density, attempts have been made to use MTJs as a Random Number Generator, exploiting its probabilistic switching characteristic. In a recent work [4], MTJs were proposed as an SNG that could produce bit streams representing any fraction between 0 and 1.

In this paper, we analyze the expected energy consumption during the switching of an MTJ as a function of the switching probability and the data being processed. Using this, we exploit the properties of stochastic numbers and MTJs to propose a simple design for a low-power MTJ-based SNG that involves data-dependent optimizations. We go on to demonstrate how it can be used to perform the same computations with a small overhead in circuit complexity but significant reductions in energy consumption.

## CHARACTERISTICS OF THE MTJ

An MTJ can exist in one of 2 states depending on the relative magnetizations of its free and fixed layers – Parallel (P, logic 0) and Anti-parallel (AP, logic 1). Depending on the switching pulse width, MTJs exhibit 3 switching modes – precessional (<3 ns), dynamic reversal (3 to 10 ns) and thermal activation (>10ns) [3]. Since we desire a high speed SNG, we operate in the precessional mode where the probability density function of the switching time is given as[1]

$$P(t_p) \propto \exp(-\Delta \sin^2\phi)(J - J_{c0})\sin^2\phi$$

with $\Delta = \frac{H_K M_s V}{2 k_B T}$ and $\phi = \frac{\pi}{2} \exp\left(-\frac{\eta \mu_B}{e M_s t_F}(J - J_{c0})t_p\right)$

We have simulated the behavior of an MTJ with in-plane magnetic anisotropy using an MTJ Spice Model[2] [5]. The values of $J_{c0}$ obtained were 7.55 MA/cm$^2$ for P->AP switching and 4.10 MA/cm$^2$ for AP->P switching and of $\Delta$ was 47.5 (sufficient retention for our purpose).

Given a pulse of width $T_p$, the probability that switching takes place within $T_p$ is $P_{sw}(T_p) = \frac{\int_0^{T_p} P(t)dt}{\int_0^{\infty} P(t)dt}$ where $P_{sw}(t)$ is the normalized cumulative distribution function (the switching probability) and the expected time at which switching takes place, given it does, is expressed as $Ex(t_{sw}) = \frac{\int_0^{T_p} t P(t) dt}{\int_0^{T_p} P(t) dt}$. Let $I_{AP}$ and $I_P$ denote the currents in the AP and P state respectively. The expected energy consumed in such a scenario, for AP->P switching, is $E_{sw}^{AP \to P}$ = V($I_{AP} Ex(t_{sw}) + I_P(T_p - Ex(t_{sw}))$), whereas the energy spent in the case where switching does not take place is $E_{nsw}^{AP \to P}$ = $V I_{AP} T_p$. Thus the overall expected energy consumed is given as Ex(E) = $P_{sw}(T_p) E_{sw}^{AP \to P}$ + (1-$P_{sw}(T_p))E_{nsw}^{AP \to P}$.

## MTJ AS A STOCHASTIC NUMBER GENERATOR

Conventionally, stochastic numbers represent probabilities. An SN with value $p$ (0 <= $p$ <= 1) is represented as a sequence of bits, such that if there are n bits in the sequence, out of which k are '1', then $p$ = k/n [2]. An MTJ can be used as an SNG by exploiting the probabilistic nature of its switching. Given a voltage pulse, the probability of switching can be decided by controlling the pulse width. The probabilities for AP->P switching, for different voltage bias, are shown in fig. 1(a).

For each bit generated by the MTJ representing a stochastic number, 3 operations need to be performed – reset the MTJ to some state, write the opposite state with some probability and read the value stored in the MTJ. Typically, one would first reset to '0' with 100% probability, write '1' with some probability $p$ and finally read the value stored in the MTJ (which would be '1' with probability $p$ and '0' with probability 1-$p$). Repeating this procedure n times would give us a sequence of n bits, out of which $p$n are expected to be 1, thereby representing the SN $p$. However, the expected energy required for switching P->AP (logic 0->1) with 99.9% probability is 0.46 pJ with a voltage bias of -0.8 V; whereas that for AP->P (logic 1->0) is 0.93 pJ with 1.2 V. We thus choose the AP state (logic 1) to be the reset state, and switch to the P state (logic 0) with some probability (because resetting from P->AP would require lesser energy than from AP->P). This means that switching AP->P with probability x will generate bit streams where the probability of finding '1' is 1-x. Hence, to represent the stochastic number $p$, we shall write 0 with a probability 1-$p$.

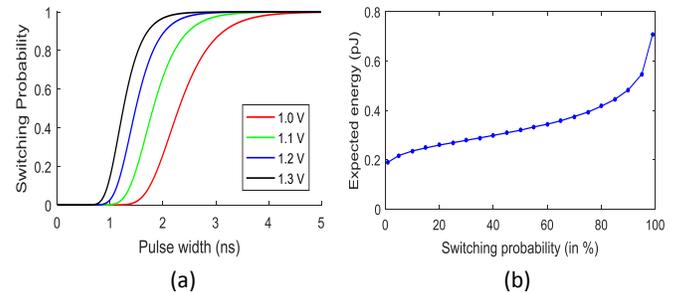

(a) (b)

Fig. 1. (a) Switching probability as a function of pulse width. (b) Expected energy v/s switching probability (AP->P switching, $V_{bias}$ = 1.2 V).

The expected energy consumption was found to be minimum for a voltage bias of about 1.2 V. We thus use this for writing to the MTJ. Shown in Fig. 1(b) is the trade-off between energy and switching probability. As stated above, resetting the MTJ is done by writing a '1' with a bias voltage of -0.8 V and pulse width of 4.33 ns. The time spent on writing '0' will depend on the switching probability desired. Switching with 99% probability (which we take as the worst case) requires 2.73 ns. Reading the value stored in the MTJ using a sense amplifier can be done with a bias of -0.1 V for 2 ns. Thus, the total time taken for generating one bit of the SN is 9.06 ns.

---

[1] $H_K$ is the shape anisotropy field, $M_s$ is the saturation magnetization, V is the volume of the free layer, $k_B$ is the Boltzmann constant, T is the temperature, J is the current density, $J_{c0}$ is the critical current density, η is the spin transfer efficiency, $t_F$ is the thickness of the free layer and $t_p$ is the pulse duration.

[2] The parameters used were - cell dimension 20 X 58 nm$^2$, $t_F$ = 2.5 nm, $M_s$ = 1222 emu/cc, α (damping constant) = 6.82 X 10$^{-3}$, η= 0.85, RA product = 5 Ω um$^2$, T = 300K.

We, however, make a slight modification to the overall procedure of generating the bits of the SN. For each bit produced, we reset the MTJ only if it was written successfully. This **Smart Reset** (SR) technique can be implemented just by using the value read from the MTJ to choose between $V_{reset}$ and 0 V (the latter indicating no reset required) using a 2:1 Mux. Further, generating an SN with value $p$, which is close to 0, with our MTJ implies that the probability of switching from AP->P (which is the same as writing '0') has to be 1-$p$, which is close to 1. Thus, more time, and hence more energy, has to be spent in writing '0' to the MTJ, as compared to the case where we had to generate an SN with value 1-$p$. To prevent this characteristic from making the SNG energy-intensive, we choose to generate 1-$p$ whenever $p$ < 0.5. In other words, instead of switching AP->P with probability 1-$p$ (which is more than 0.5), we switch with probability $p$ (which is less than 0.5). Now all we need to do is to invert the bits output from this **Biased MTJ-SNG** (BMS, the name being derived from the biased nature of the data produced by the MTJ-SNG) so that we get the SN $p$. We may also be able to use them directly for certain computations. Note that this technique of generating 1-$p$ instead of $p$ is used only when $p$ < 0.5; for $p$ > 0.5, the process remains as described earlier. We illustrate the benefits of these techniques through an example, multiplication of 2 stochastic numbers, in the next section.

The energy required to generate one bit from the MTJ-SNG is plotted in Fig. 2 as a function of the switching probability for all the 3 cases - without any modification (Normal), with Smart Reset and finally with both SR and the BMS in place. The symmetry of the plot with Biased MTJ-SNG comes from choosing between the larger of $p$ and 1-$p$. These techniques let us achieve a minimum energy saving of 29% (which occurs with $p$ = 0.5), a maximum of 83% (with $p$ = 0.99) and an average of 54% for each bit produced by the MTJ. Since the BMS requires us to generate stochastic numbers only greater than or equal to 0.5, the maximum write duration reduces from 2.73 ns to 1.49 ns (which corresponds to the pulse width giving 50% switching probability).

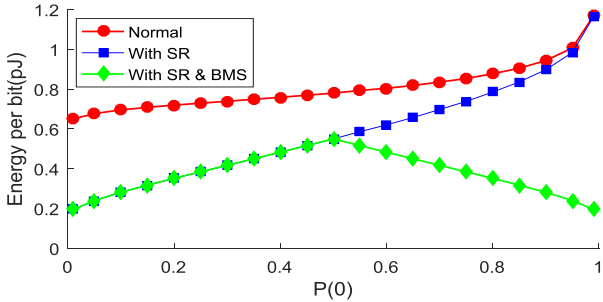

Fig. 2. Energy required to generate one bit from the MTJ-SNG as a function of the probability of having a 0 ($p$ = 1 - P(0)) for the 3 arrangements.

## MULTIPLICATION OF STOCHASTIC NUMBERS

To multiply two SNs $p_1$ and $p_2$, we would typically generate them as they are and use an AND gate to obtain the product [2]. However, as we saw in the last section, if an SN $p$ is less than 0.5, then generating 1-$p$ and inverting the bits is more energy efficient than generating $p$ itself. That is why, when any of the SNs to be multiplied (say $p$) is less than 0.5, we write to its MTJ with a probability $p$, thereby generating the SN (1-$p$). This results in the multiplier circuit which is explained below and depicted in Fig. 3.

For multiplying $p_1$ and $p_2$, we generate either $p_1$ or 1-$p_1$ (and either $p_2$ or 1-$p_2$), whichever is larger and use a 2:1 multiplexer to choose between the two. The inputs to the Select pins of the multiplexers can be derived from the most significant bit of the binary number that is being converted to a stochastic number [2]. Thus we can always obtain the product $p_1p_2$ by generating the larger of $p_1$ and 1-$p_1$ and also of $p_2$ and 1-$p_2$ and selecting accordingly between themselves and their inverse.

For example, if $p_1$ = 0.7 and $p_2$ = 0.2, the first MTJ will generate $p_1$ (so A = 0.7), but the second one will generate 1-$p_2$ (so B = 0.8). $S_X$ will be 0 to choose A and $S_Y$ will be 1 to choose B'. Thus, X = 0.7 and Y = 0.2, thereby making Z = 0.14 which is $p_1p_2$. As per calculations, using SR and BMS reduces energy consumption by about 45 % from 1.95 pJ to 1.07 pJ. Since multiplication is one of the most fundamental operations in stochastic computation, with numerous applications, such as in polynomial function synthesis [4], implementation using these 2 techniques (SR and BMS) is likely to save a lot of energy.

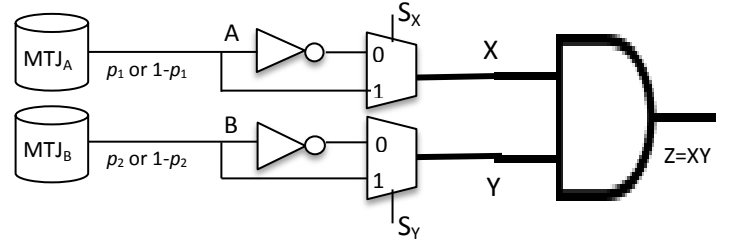

Fig. 3. Proposed multiplier circuit to be used with MTJ-SNGs. X = A'$S_X$ + AS$_X$' and Y = B'$S_Y$ + BS$_Y$'. A and B are always greater than or equal to 0.5.

Table 1 provides the energy and power consumption of the MTJ-SNG and the stochastic multiplier (consisting of 2 SNGs and the multiplier circuit) for all the 3 cases. The energy required per bit (2nd row) covers the range of values of $p$ from 0.01 to 0.99. Note, however, that the energy required is least when $p$ = 0.99 for Normal and SR and when $p$ = 0.99 or 0.01 for SR & BMS. Whereas it is most when $p$ = 0.01 for Normal and SR and when $p$ = 0.5 for SR & BMS. The average energy and power have been calculated considering a uniform distribution of $p$ over all probability values in the range [0, 1]. Table 2 gives an idea of how the energy required for multiplication varies with the precision of the SNs.

|  | Normal | Smart Reset | SR & BMS |
|---|---|---|---|
| Time per bit (ns) | 9.06 | 9.06 | 7.82 |
| Energy per bit (pJ) | 0.65-1.16 | 0.19-1.16 | 0.19-0.55 |
| Avg. Energy per bit (pJ) | 0.809 | 0.579 | 0.372 |
| MTJ-SNG Avg. Power (mW) | 0.107 | 0.083 | 0.067 |
| Multiplier Avg. Power (mW) | 0.215 | 0.166 | 0.135 |
| Multiplier Avg. Energy (pJ) | 1.95 | 1.50 | 1.06 |

Table 1. Energy and power estimates with the MTJ-SNG and the multiplier

| Precision | No. of bits | Energy (nJ) | | |
|---|---|---|---|---|
|  |  | Normal | Smart Reset | SR and BMS |
| 8-bit | 256 | 0.50 | 0.38 | 0.27 |
| 9-bit | 512 | 1.00 | 0.77 | 0.54 |
| 10-bit | 1024 | 2.00 | 1.54 | 1.09 |
| 11-bit | 2048 | 3.99 | 3.07 | 2.17 |
| 12-bit | 4096 | 7.99 | 6.14 | 4.34 |

Table 2. Energy spent in multiplying 2 SNs with different levels of precision

In summary, we have illustrated the probabilistic nature of MTJ switching and the trade-off between energy consumption and switching probability. An energy efficient design of an MTJ as an SNG is proposed that can be used for all stochastic computations. Its benefits are demonstrated with a multiplier circuit that allows us to reduce computation time by 14% and average energy by 46% when compared to the multiplier with the baseline MTJ-SNG.